\documentclass[preprint,showpacs,preprintnumbers,amsmath,amssymb]{revtex4}
\usepackage[dvips]{epsfig}
\def\be{\begin{equation}}
\def\ee{\end{equation}}
\def\bdm{\begin{eqnarray}}
\def\edm{\end{eqnarray}}
\begin{document}
\preprint{Submitted to Physics of Plasmas}
\title{Random walk of magnetic field lines for different values of the energy-range spectral index}
\author{A. Shalchi \& I. Kourakis}
\affiliation{Institut f\"ur Theoretische Physik, Lehrstuhl IV:
Weltraum- und Astrophysik, Ruhr-Universit\"at Bochum, D-44780
Bochum, Germany}
\date{\today}
\begin{abstract}
An analytical nonlinear description of field-line wandering in partially statistically magnetic systems was proposed recently. 
In this article we investigate the influence of the wave-spectrum in the energy-range onto field line random walk by applying
this formulation. It is demonstrated that in all considered cases we clearly obtain a superdiffusive behaviour of the field-lines. 
If the energy-range spectral index exceeds unity a free-streaming behaviour of the field-lines can be found for all relevant 
length-scales of turbulence. Since the superdiffusive results obtained for the slab model are exact, it seems that superdiffusion 
is the normal behavior of field line wandering.
\end{abstract}
 \maketitle
\section{Introduction}
Understanding turbulence is an issue of major importance in space physics and astrophysics; see, e.g., in Refs. 
\cite{MC,RS,gol95,cho02,zho04,Isio92}. It has been demonstrated in several articles that stochastic wandering of magnetic 
field-lines directly influences the transport of charged cosmic rays (see e.g. \cite{Jo73,Skill74,Nar01,Matt03,Chan04,Maron04,
kot00,web06,ShaKo07a}. Several theories have been developed to describe field-line random-walk (FLRW) analytically. The 
classic work of Jokipii (see \cite{jok66}), for instance, employed a quasilinear approach for FLRW. In this theory the unperturbed 
field-lines are used to describe field-line wandering by using a perturbation method. It has often been stated that this approach is 
correct in the limit of weak turbulence where it is assumed that the turbulent fields are much weaker than the uniform mean field 
($\delta B_i \ll B_0$). To achieve a more reliable and general description of field-line wandering Matthaeus \textit{et al.} (see 
\cite{mat95}) developed a nonperturbative statistical approach by combining certain assumptions about the properties of the field-lines 
(e.g. Gaussian statistics) with a diffusion model. More precisely, in the Matthaeus et al. theory of field-line wandering is has explicitly 
been assumed that field-line wandering behaves diffusively.

An improved theory for FLRW, which is essentially a generalization of the theory of Matthaeus \textit{et al.}, was recently developed
by Shalchi \& Kourakis (see \cite{ShaKo07b}). By explicitly assuming a diffusive behavior of the field-lines, the Matthaeus \textit{et al.} 
theory can be obtained from the Shalchi \& Kourakis approach as a special limit. However, it has also been demonstrated in
\cite{ShaKo07b} that for slab/2D turbulence geometry, the field-lines behave superdiffusively. Thus, the Matthaeus et al. theory 
cannot be applied for slab/2D composite geometry. As also demonstrated in \cite {ShaKo07b}, quasilinear theory is only correct for 
pure slab geometry or for small length scales.

In most past studies a constant spectrum in the energy-range has been assumed (in this case the energy-range spectral index is
equal to zero). It is the purpose of this article to explore different values of the energy-range spectral index. The layout of
this article goes as follows. In Section 2, we discuss different forms of the wave-spectrum which are appropriate for solar wind
turbulence. In Section 3, we calculate the FLRW for pure slab geometry for different values of the energy-range spectral index
by applying the exact formulation for field-line wandering. In Section 4, we employ the nonlinear theory of Shalchi \& Kourakis
for FLRW, in order to deduce an analytic form for the field-line MSD for pure 2D turbulence. These results can easily be
combined with the pure 2D result to describe field-line wandering in the slab/2D composite model (Section 5). In Section 6 we
summerize our new results.
\section{Different forms of the wave-spectrum}
In \cite{bie96} a two-component turbulence model has been proposed as a realistic model for solar wind turbulence. In this
model we describe the turbulent fields as a superposition of a slab model ($\vec{k} \parallel \vec{B}_0$) and a 2D model
($\vec{k} \perp \vec{B}_0$). In this case the $xx-$component of the correlation tensor can be written as 
\be 
P_{xx} (\vec{k}) = P_{xx}^{slab} (\vec{k}) + P_{xx}^{2D} (\vec{k}) 
\label{xxcorr} 
\ee
with the slab contribution 
\be 
P_{xx}^{slab} (\vec{k}) = g^{slab}(k_{\parallel}) {\delta (k_{\perp}) \over k_{\perp}}
\label{xxslab} 
\ee 
and the 2D contribution 
\be 
P_{xx}^{2D} (\vec{k}) = g^{2D} (k_{\perp}) {\delta (k_{\parallel}) \over k_{\perp}} \left[ 1 - {k_x^2 \over k^2} \right]. 
\label{xx2d} 
\ee
In previous studies the forms 
\be 
g^{slab} (k_{\parallel}) = {C(\nu) \over 2 \pi} l_{slab} \delta B_{slab}^2 (1 + k_{\parallel}^2 l_{slab}^2)^{-\nu} 
\label{oldslabspec} 
\ee 
for the slab wave-spectrum, and 
\be 
g^{2D} (k_{\perp}) = {2 C(\nu) \over \pi} l_{2D} \delta B_{2D}^2 \, (1 + k_{\perp}^2 l_{2D}^2)^{-\nu}
\label{old2dspec} 
\ee 
for the 2D wave-spectrum were used. The \emph{energy-range} of the spectrum is defined for $k_{\parallel} \ll l_{slab}^{-1}$ and 
$k_{\perp} \ll l_{2D}^{-1}$. Clearly, both spectrum forms are constant in the energy-range. However, as discussed in several previous 
articles (see e.g. \cite{bru05}), we find in heliospheric observations a steeper spectrum (according to \cite{bru05} the energy-range 
spectral index - cf. (\ref{slabspec}) below - should be $q=1.07$). In the following we deduce and discuss analytical forms of the wave
spectrum for slab and 2D turbulence models.
\subsection{General form of the slab wavespectrum}
According to solar wind observations, the following form of the spectrum should be appropriate: 
\bdm 
g^{slab}(k_{\parallel}) & = & \frac{c_i}{2 \pi} l_{slab} \delta B_{slab}^2 \nonumber\\
& \times & \!\left\{\!\begin{array}{ccc}
0 & \textnormal{if } & k_{\parallel}<k_{\min}\\
(k_{\parallel}\ l_{slab})^{-q} & \textnormal{if } & k_{\min}\leq k_{\parallel} \leq l_{slab}^{-1}\\
(k_{\parallel}\ l_{slab})^{-2v} & \textnormal{if} & l_{slab}^{-1}<k_{\parallel}.
\end{array}\right.
\label{slabspec} 
\edm 
Here we defined the slab-bendover-scale $l_{slab}$, the strength of the turbulent field $\delta B_{slab}^2$, and the inertial-range spectral index 
$2 \nu$. $k_{min}$ indicates the smallest wave-number which might be related to the bulk plasma length scale $L$ via $k_{min} \sim L^{-1}$. We 
have also introduced the energy-range spectral index $q$. By taking into account the normalization condition 
\be 
\delta B_{slab}^2 = \delta B_{x}^2 + \delta B_{y}^2 = \int d^3 k \; \left[ P_{xx}^{slab} (\vec{k}) + P_{yy}^{slab} (\vec{k}) \right]
\label{slabnorm} 
\ee 
we find for the normalization constant $c_i$ the values shown in Table \ref{slabtab}. The values shown there are valid if the condition 
$k_{min} l_{slab} \ll 1$ is fulfilled.
\begin{table}[t]
\begin{center}
\begin{tabular}{|l|l|l|}\hline
\vphantom{$1 \over 2$} $ \textnormal{case}	$ & $ \textnormal{Normalization constants $c_i$}                                    			$ \\
\hline\hline
\vphantom{$1 \over 2$} $    0 < q < 1               $ & $ c_1:=\left(\frac{4}{1-q}+\frac{4}{2v-1}\right)^{-1}                           			$ \\
\vphantom{$1 \over 2$} $    q = 1                      $ & $ c_2:=\frac{1}{4}\left(\ln\!\Bigl({\frac{1}{k_{\min}\ l_{slab}}}\Bigr)\right)^{-1} 	$ \\
\vphantom{$1 \over 2$} $    1 < q < 2               $ & $ c_3:=\frac{q-1}{4} \Bigl(k_{\min}\ l_{slab}\Bigr)^{q-1}                           		$ \\
\hline
\end{tabular}
\caption{\label{slabtab} The exact values of the various normalization constants $c_i$ ($i=1, 2, 3$) are provided. These expressions are 
correct for $k_{min} l_{slab} \ll 1$.}
\end{center}
\end{table}
\subsection{General form of the 2D wavespectrum}
For the 2D spectrum we can adopt the same form for the spectrum as used in the last subsection for the slab spectrum:
\bdm
g^{2D}(k_{\perp}) & = & \frac{d_i}{2 \pi} l_{2D} \delta B_{2D}^2
\nonumber\\
& \times & \!\left\{\!\begin{array}{ccc}
0 & \textnormal{if } & k_{\perp}<k_{\min}\\
(k_{\perp}\ l_{2D})^{-q} & \textnormal{if } & k_{\min}\leq k_{\perp} \leq l_{2D}^{-1}\\
(k_{\perp}\ l_{2D})^{-2v} & \textnormal{if} & l_{2D}^{-1}<k_{\perp}.
\end{array}\right.
\label{2dspec} 
\edm 
Here we used the 2D-bendover-scale $l_{2D}$, the strength of the turbulent field $\delta B_{2D}^2$, and the inertial-range spectral index $2 \nu$. 
$k_{min}$ indicates again the smallest wave-number and $q$ is again the energy-range spectral index. Fulfilling the normalization condition 
\be 
\delta B_{2D}^2 = \delta B_{x}^2 + \delta B_{y}^2 = \int d^3 k \; \left[ P_{xx}^{2D} (\vec{k}) + P_{yy}^{2D} (\vec{k}) \right]
\label{2dnorm} 
\ee 
we find for the normalization constant $d_i$ the values shown in table \ref{2dtab}.
\begin{table}[t]
\begin{center}
\begin{tabular}{|l|l|l|}\hline
\vphantom{$1 \over 2$} $ \textnormal{case}	$ & $ \textnormal{Normalization constants $d_i$}                            $ \\
\hline\hline
\vphantom{$1 \over 2$} $    0 < q < 1               $ & $ d_1:=\left(\frac{1}{1-q}+\frac{1}{2v-1}\right)^{-1}                           $ \\
\vphantom{$1 \over 2$} $    q = 1                   	$ & $ d_2:=\left(\ln\!\Bigl({\frac{1}{k_{\min}\ l_{2D}}}\Bigr)\right)^{-1}              $ \\
\vphantom{$1 \over 2$} $    1 < q < 2               $ & $ d_3:= (q-1) \Bigl(k_{\min}\ l_{2D}\Bigr)^{q-1}                            $ \\
\hline
\end{tabular}
\caption{\label{2dtab} The exact values of the various normalization constants $d_i$ ($i=1, 2, 3$) are provided. These
expressions are correct for  $k_{min} l_{2D} \ll 1$.}
\end{center}
\end{table}
In the following we consider different values of the energy-range spectral index $q$ and calculate the field-line mean square deviation
analytically for pure-slab, pure-2D, and two-component turbulence. 
\section{FLRW for pure-slab turbulence}
As shown in several previous papers (e.g. \cite{ShaKo07b}) the field-line mean square deviation can be calculated 
exactly for pure slab geometry. For standard forms of the wave spectrum, where $q=0$, we find the classical diffusive result: 
$\left< \left( \Delta x \right)^2 \right> = 2 \kappa_{FL} | z |$ (see e.g. \cite{jok66,mat95}), with the field-line diffusion 
coefficient $\kappa_{FL}$. Shalchi \& Kourakis (see \cite{ShaKo07b}) derived the following ordinary differential equation (ODE) 
for the mean square deviation and slab geometry
\bdm
{d^2 \over d z^2} \left< \left( \Delta x \right)^2 \right> & = & {2 \over B_0^2} \int d^3 k \; P_{xx}^{slab} (\vec{k}) \cos (k_{\parallel} z) \nonumber\\
& = & {8 \pi \over B_0^2} \int_{0}^{\infty} d k_{\parallel} \; g^{slab} (k_{\parallel}) \cos (k_{\parallel} z). 
\label{slab1}
\edm 
For the wave spectrum of Eq. (\ref{slabspec}) this becomes
\bdm 
{d^2 \over d z^2} \left< \left( \Delta x \right)^2 \right> & \approx & 4 c_i l_{slab}^{1-q}{\delta B_{slab}^2 \over B_0^2}
\int_{k_{min}}^{l_{slab}^{-1}} d k_{\parallel} \; k_{\parallel}^{-q} \cos (k_{\parallel} z) + \dots 
\label{slab2}
\edm 
It can easily be proven that the contribution of the inertial-range ($k_{\parallel} \geq l_{slab}^{-1}$) is much smaller than the contribution of the 
energy-range ($k_{\parallel} \leq l_{slab}^{-1}$), and was thus neglected in the right-hand side (rhs) of Eq. (\ref{slab2}). Furthermore, the upper 
limit of the $k_{\parallel}$-integral can be extended to infinity. Here $q>0$ is assumed, for convergence.

Taking into account the relation 
\be 
\int_{u}^{\infty} d x \; x^{\mu-1} \cos x = {1 \over 2} \left[ e^{- i \pi \mu / 2} \Gamma \left( \mu, +i u \right) + e^{+ i \pi \mu / 2} \Gamma \left( \mu,
-i u \right) \right]   \, , 
\label{slab3} 
\ee 
according to Gradshteyn \& Ryzhik (see \cite{Grad}, page 430, Eq. 3.761.7 therein), for $\mu<1$ (implying here $q > 0$), where we have 
employed the incomplete Gamma function $\Gamma (\mu, x) = \int_{x}^{\infty} dt \; t^{\mu-1} e^{-t}$ (see Eq. 8.35 in the latter reference), and 
approximating $\Gamma(\mu, x)$, for small values of the argument $x$, as
\be 
\Gamma ( \mu, x \ll 1 ) \approx \Gamma \left(  \mu \right) \left[ 1 - {x^\mu \over  \mu \Gamma (  \mu )} \right]  
\label{slab4} 
\ee
(see Eq. 8.354.2 in the same reference) we find 
\be 
\int_{u}^{\infty} d x \; x^{\mu-1} \cos x \approx \Gamma \left( \mu \right) \cos \left( {\pi \over 2} \mu \right) - {1 \over \mu} u^{\mu}. 
\label{slab5}
\ee
By applying this formula onto Eq. (\ref{slab2}) one gets 
\bdm
{d^2 \over d z^2} \left< \left( \Delta x \right)^2 \right> & = & 4 c_i l_{slab}^{1-q} {\delta B_{slab}^2 \over B_0^2} \, z^{q-1} \nonumber\\
& \times & \left[ \Gamma \left( 1-q \right) \sin \left( {\pi q \over 2} \right) + {1 \over q-1} (z k_{min})^{1-q} \right].
\label{slab6}
\edm
The result can easily be integrated to obtain
\bdm
\left< \left( \Delta x \right)^2 \right> & = & {4 c_i \over q (q+1)} l_{slab}^{1-q} {\delta B_{slab}^2 \over B_0^2}\, z^{q+1} \nonumber\\
& \times & \left[ \Gamma \left( 1-q \right) \sin \left( {\pi q \over 2} \right) + {q (q+1) \over 2 (q-1)} (z k_{min})^{1-q} \right]. 
\label{slab7} 
\edm 
This expression is valid for $0 < q < 1$ and for $1 < q < 2$. For $q=1$, Eq. (\ref{slab2}) can be directly evaluated and we find a logarithmic 
behavior of the MSD. In the following, we shall further simplify Eq. (\ref{slab7}), by distinguishing the ranges $0 < q < 1$ and $1 < q < 2$. We
stress that the are interested in the large $z$ range, although we note that the condition $z k_{min} = \epsilon \ll 1$ is assumed to hold 
everywhere (since $k_{min}$ is related to the inverse size of the plasma ``box''). We therefore retain the definition of the small parameter 
$\epsilon$, whose polynomial contribution may be singled out, for order of magnitude estimates.
\subsection{Smooth spectrum form: the case $0 < q < 1$}
In this case the first term in the rhs of Eq. (\ref{slab7}) is dominant and we obtain 
\bdm 
\left< \left( \Delta x \right)^2 \right> \approx {4 c_1 \over q (q+1)} l_{slab}^{1-q} z^{q+1} {\delta B_{slab}^2 \over B_0^2} \Gamma \left( 1-q \right) \sin
\left( {\pi q \over 2} \right) \sim z^{q+1} \, ,
\label{slab8} 
\edm
while a contribution $\sim {\cal O}(\epsilon^{1-q})$ within the brackets in (\ref{slab7}) is omitted. In general the mean square deviation of the 
field-lines has the form $\left< \left( \Delta x \right)^2 \right> = a z^{b}$. According to Eq. (\ref{slab8}) we find for the slab model and for the 
values of the energy-range spectral index considered the characteristic exponent 
\be 
b= q+1.
\label{slab8a} 
\ee
It is obvious that we obtain superdiffusion ($1 < b < 2$) for $0 < q < 1$.
\subsection{Steep spectrum form: the case $1 < q < 2$}
In this case the second term in the rhs of Eq. (\ref{slab7}) is dominant (of the order $\epsilon^{1-q} \gg 1$) and one gets 
\be
\left< \left( \Delta x \right)^2 \right> = {z^2 \over 2} {\delta B_{slab}^2 \over B_0^2}. 
\label{slab10} 
\ee 
This result if formally the same as the initial free-streaming result which can be found for small $z$ values (see e.g. \cite{ShaKo07b}).
\section{FLRW for pure 2D turbulence}
In this Section, we shall follow the nonlinear formalism for FLRW proposed by Shalchi \& Kourakis (\cite{ShaKo07b}). 
According to the results therein we have for pure 2D turbulence 
\be 
{d^2 \over d z^2} \left< \left( \Delta x \right)^2 \right> = {2 \pi \over B_0^2} \int_{0}^{\infty} d k_{\perp} \; g^{2D} (k_{\perp}) e^{-{1 \over
2} \left< \left( \Delta x \right)^2 \right> k_{\perp}^2}.
\label{2d1} 
\ee 
With the spectrum of Eq. (\ref{2dspec}) we find
\be 
{d^2 \over d z^2} \left< \left( \Delta x \right)^2 \right> \approx d_i l_{2D}^{1-q} {\delta B_{2D}^2 \over B_0^2}
\int_{k_{min}}^{l_{2D}^{-1}} d k_{\perp} \; k_{\perp}^{-q} e^{-{1\over 2} \left< \left( \Delta x \right)^2 \right> k_{\perp}^2} + \dots 
\label{2d2} 
\ee 
The detailed calculation (limited to the case $q=0$) was carried out in the latter reference, where a superdiffusive behavior of the 
form $\left< \left( \Delta x \right)^2 \right> \sim z^{4/3}$ was obtained. Our purpose in the following is to extend that result, for a general 
form of the wave spectrum.

It can easily be demonstrated that the inertial-range of the spectrum yields a negligible contribution in the rhs of (\ref{2d2}) and 
was thus here neglected. The integral from the energy-range, extending the upper limit to infinity ($l_{2D}^{-1} \rightarrow \infty$), 
can be expressed by Gamma functions 
\bdm 
{d^2 \over d z^2} \left< \left( \Delta x \right)^2 \right> & = & {d_i \over 2} l_{2D}^{1-q} {\delta B_{2D}^2 \over B_0^2}
\left( {\left< (\Delta x)^2 \right> \over 2} \right)^{(q-1)/2} \nonumber\\
& \times & \left[ \Gamma \left( {1-q \over 2} \right) + \Gamma \left( {1-q \over 2}, {1 \over 2} \left( \left< (\Delta x)^2
\right> k_{min}^2 \right)^2 \right) \right] \, . 
\label{2d3} 
\edm

Assuming that $\left< (\Delta x)^2 \right> k_{min}^2 \ll 1$ (i.e., the field-line MSD cannot exceed the maximum turbulence square
length scale $k_{min}^{-2}$), and using Eq. (\ref{slab4}) we find
\bdm 
{d^2 \over d z^2} \left< \left( \Delta x \right)^2 \right> & \approx & d_i l_{2D}^{1-q} {\delta B_{2D}^2 \over B_0^2}
\left( {\left< (\Delta x)^2 \right> \over 2} \right)^{(q-1)/2} \nonumber\\
& \times & \left[ \Gamma \left( {1-q \over 2} \right) + {1 \over q-1} \left( {\left< (\Delta x)^2 \right> k_{min}^2 \over 2}
\right)^{(1-q)/2} \right]. 
\label{2d5} 
\edm 
The formula can be applied so long as $0 < q < 2$, except for $q=1$. In the latter case, Eq. (\ref{2d2}) can be directly evaluated and we find a
logarithmic behavior of the MSD. In the following, we shall further simplify Eq. (\ref{2d5}), separately considering the cases $0 \leq q < 1$ and 
for $1 < q < 2$. The relation $\epsilon' =\left< (\Delta x)^2 \right> k_{min}^2 \ll 1$ is assumed to hold everywhere.
\subsection{The case $0 < q < 1$}
In this case the first term in Eq. (\ref{2d5}) is dominant 
\be
{d^2 \over d z^2} \left< \left( \Delta x \right)^2 \right> \approx d_1 l_{2D}^{1-q} {\delta B_{2D}^2 \over B_0^2} \left( {\left<
(\Delta x)^2 \right> \over 2} \right)^{(q-1)/2} \Gamma \left( {1-q \over 2} \right). 
\label{2d6} 
\ee 
By making the \emph{ansatz} $\left< \left( \Delta x \right)^2 \right> = a \, z^{b}$ we can solve this ODE analytically. It can easily be demonstrated that
\be 
b= {4 \over 3 - q} \label{2d7} \ee and \be a = \left[ d_1 l_{2D}^{1-q} {\delta B_{2D}^2 \over B_0^2} 2^{(1-q)/2} \, \frac{(3-q)^2}{4 (1+q)} \, 
\Gamma \left( {1-q \over 2} \right) \right]^{2/(3-q)}. \label{2d8} \ee Obviously we find \be {4 \over 3} < b < 2 
\label{2d9} 
\ee 
which is interpreted as \emph{superdiffusion}. A diffusive behavior ($b=1$) cannot be obtained. Interestingly, even for $q=0$, one finds $b=4/3$ 
(see in \cite{ShaKo07b}).
\subsection{The case $1 < q < 2$}
In this case the second term within brackets in Eq. (\ref{2d5}) is dominant (of the order $\sim \epsilon^{(1-q)/2} \gg 1$; see above)
and we have 
\be 
{d^2 \over d z^2} \left< \left( \Delta x \right)^2 \right> = {d_3 \over q-1} {\delta B_{2D}^2 \over B_0^2} \left(l_{2D} k_{min} \right)^{1-q}. 
\label{2d10} 
\ee 
By using Table \ref{2dtab} for $d_3$ this can be simplified to 
\be 
{d^2 \over d z^2} \left< \left( \Delta x \right)^2 \right> = {\delta B_{2D}^2 \over B_0^2} 
\label{2d11} 
\ee 
and we finally find 
\be 
\left< \left( \Delta x \right)^2 \right> = {z^2 \over 2} {\delta B_{2D}^2 \over B_0^2} 
\label{2d12} 
\ee 
which is again the initial free-streaming (parabolic MSD) result.
\section{FLRW for slab/2D composite geometry}
According to cosmic observations, it is more realistic than plainly adopting a pure-slab or pure-2D model, to consider a 20\% slab/80\% 2D 
composite model (see e.g. \cite{bie96}). In this case, one rigorously obtains a 2nd-order ODE [cf. (\ref{slab1}), (\ref{2d1})], whose RHS is the 
sum of the slab and 2D contributions, the relative weight of which is determined by the corresponding turbulence strength, i.e., 
$\delta B_{slab}^2/\delta B^2$ and $\delta B_{2D}^2/\delta B^2$.

We shall now attempt to evaluate the asymptotic behavior of the field-line MSD in this hybrid (composite) model.
\subsection{The case $0 < q < 1$}
In this case we can combine Eqs. (\ref{slab6}) and (\ref{2d6}) into: 
\bdm 
{d^2 \over d z^2} \left< \left( \Delta x \right)^2 \right> & = & 4 c_1 l_{slab}^{1-q} z^{q-1} {\delta B_{slab}^2 \over B_0^2}
\Gamma \left( 1-q \right) \sin \left( {\pi q \over 2} \right) \nonumber\\
& + & d_1 l_{2D}^{1-q} {\delta B_{2D}^2 \over B_0^2} \left( {\left< (\Delta x)^2 \right> \over 2} \right)^{(q-1)/2} \Gamma
\left( {1-q \over 2} \right) \, ,  
\edm 
where negligible contributions were omitted in the rhs. Obviously this equation has the form 
\be 
{d^2 \over d z^2} \left< \left( \Delta x \right)^2 \right> = \alpha z^{q-1} + \beta \left[ \left< (\Delta x)^2 \right> \right]^{(q-1)/2}. 
\ee 
By applying the \emph{ansatz} $\left< (\Delta x)^2 \right> = a z^{b}$ we find 
\be 
a b (b-1) z^{b-2} = \alpha z^{q-1} + \beta a^{(q-1)/2} z^{b(q-1)/2} \, 
\ee
where the definitions of $\alpha$ and $\beta$ are obvious. It is straightforward to prove that, since $b < 2$, the second term in the rhs is 
dominant for $z \rightarrow \infty$. The slab contribution can therefore be neglected, so we can use Eqs. (\ref{2d7}) and (\ref{2d8}) also 
within the two-component model.
\subsection{The case $1 < q < 2$}
In this case we can simply add the two contributions (Eqs. (\ref{slab10}) and (\ref{2d12})): 
\be 
\left< \left( \Delta x \right)^2 \right> = {z^2 \over 2} {\delta B^2 \over B_0^2} 
\ee
where we have set 
\be 
\delta B^2 = \delta B_{slab}^2 + \delta B_{2D}^2. 
\ee
\section{Summary and conclusion}
We have investigated the random walk of magnetic field-lines for a more general spectrum, than the one employed in previous 
works. By exploring pure slab, pure 2D, and two-component turbulence models, we have calculated the field-line mean square 
deviation by applying the analytical description for FLRW proposed by Shalchi \& Kourakis (see \cite{ShaKo07b}). A superdiffusive 
behaviour is found in all cases considered. In Table \ref{resulttab} the results obtained in this article are summarized.
\begin{table}[t]
\begin{center}
\begin{tabular}{|l|l|l|l|}\hline
\vphantom{$1 \over 2$} $ \textnormal{Geometry} $ & $ \textnormal{spectral index}    $ & $ \textnormal{a}                            $ & $ \textnormal{b}    $ \\
\hline\hline \vphantom{$1 \over 2$} $ slab                   $ & $
0 < q < 1                $ & $
{4 c_1 \over q (q+1)} l_{slab}^{1-q} {\delta B_{slab}^2 \over B_0^2} \Gamma \left( 1-q \right) \sin \left( {\pi q \over 2} \right)              $ & $ q+1           $ \\
\vphantom{$1 \over 2$} $ slab                   $ & $   1 < q < 2                   $ & $ {1 \over 2} {\delta B_{slab}^2 \over B_0^2}   $ & $ 2             $ \\
\vphantom{$1 \over 2$} $ 2D                     $ & $   0 < q <
1                $ & $
\left[ d_1 l_{2D}^{1-q} {\delta B_{2D}^2 \over B_0^2} 2^{(1-q)/2} \Gamma \left( {1-q \over 2} \right) \right]^{2/(3-q)}                    $ & $ 4 / (3 - q)       $ \\
\vphantom{$1 \over 2$} $ 2D                     $ & $   1 < q < 2                   $ & ${1 \over 2} {\delta B_{2D}^2 \over B_0^2}  $ & $ 2             $ \\
\vphantom{$1 \over 2$} $ slab/2D                $ & $   0 < q < 1                $ & $
\left[ d_1 l_{2D}^{1-q} {\delta B_{2D}^2 \over B_0^2} 2^{(1-q)/2} \Gamma \left( {1-q \over 2} \right) \right]^{2/(3-q)}                    $ & $ 4 / (3 - q)       $ \\
\vphantom{$1 \over 2$} $ slab/2D                $ & $   1 < q < 2                   $ & ${1 \over 2} {\delta B^2 \over B_0^2}       $ & $ 2             $ \\
\hline
\end{tabular}
\caption{\label{resulttab} In this table, the results obtained for the parameters $a$ and $b$, having adopted the form $\left< \left(
\Delta x \right)^2 \right> = a z^{b}$ for the field-line mean square deviation, are presented. In all (but one) cases, we find
either superdiffusion ($1<q<2$) or free-streaming ($q=2$) of the field-lines. Diffusion ($q=1$) can only be found for slab geometry
and $q=0$. }
\end{center}
\end{table}
The only case where one obtains diffusion is for pure slab geometry and $q=0$. As shown in this article the energy-range
spectral index is a key-input parameter if FLRW is described.

In the two-component turbulence model, which has been considered as a realistic model for solar wind turbulence (see \cite{bie96}), we already 
find a weakly superdiffusive behavior if $q=0$. For larger values of $q$ we have $\left< \left( \Delta x \right)^2 \right> \sim z^{4/(3-q)}$. If the 
energy-range spectral index exceeds unity we find the same solution as in the initial free-streaming regime. Obviously,  the energy-range spectral 
index has a very strong influence on FLRW behavior. 

From a theoretical point of view the results for pure slab geometry deduced in Section 3 are very interesting and important 
because of two reasons:
\begin{itemize}
\item for pure slab turbulence the parameter $<(\Delta x)^2>$ can be calculated exactly. No theory nor any ad hoc assumption 
have to be applied.
\item In all cases except $q=0$ we find superdiffusion of FLRW.
\end{itemize}
Since in reality 20 \% of the fluctuations can be represented by slab modes (see \cite{bie96}) it is self-evident to assume
that superdiffusion and not (classical or Markovian) diffusion is the regular case in astrophysical turbulence.

If we merge from pure slab geometry to the slab/2D composite model a (nonlinear) theory has to be applied and an exact
description of FLRW in no longer possible. By applying the ODE deduced by Shalchi \& Kourakis (see \cite{ShaKo07b}) we have 
shown that the superdiffusivity becomes even stronger in comparison to the pure slab results.

It must be the subject of future work to apply these new results on realistic systems, such as solar wind turbulence. An important 
example is perpendicular transport of charged cosmic rays which is directly controlled by the FLRW, since charged particles are 
tied to magnetic field-lines (see \cite{ShaKo07a}).
\begin{acknowledgments}
This research was supported by Deutsche Forschungsgemeinschaft (DFG) under the Emmy-Noether Programme (grant SH 93/3-1). As a
member of the {\it Junges Kolleg} A. Shalchi also aknowledges support by the Nordrhein-Westf\"alische Akademie der Wissenschaften.
\end{acknowledgments}
{}

\begin{thebibliography}{}
%
\bibitem{MC} W.D. Mc Comb, {\it The physics of fluid turbulence} (Oxford Science Publications, UK, 1990).
\bibitem{RS} R. Schlickeiser, {\it Cosmic Ray Astrophysics} (Springer, Berlin, 2002).
\bibitem{gol95} P. Goldreich, S. Sridhar, Astrophys. J., 438, 763 (1995).
\bibitem{cho02} J. Cho, A. Lazarian, E. T. Vishniac, Astrophys. J., 564, 291(2002).
\bibitem{zho04} Y. Zhou, W. H. Matthaeus, P. Dmitruk, Rev. Mod. Phys., 76, 1015 (2004).
\bibitem{Isio92} M. B. Isichenko, Rev. Mod. Phys. 64, 961 - 1043 (1992).
\bibitem{Jo73} J. R. Jokipii, Astrophys. J., 183, 1029 (1973).
\bibitem{Skill74} J. Skilling, I. McIvor, J. Holmes, MNRAS, 167, 87P (1974).
\bibitem{Nar01} R. Narayan, R. Medvedev, M. Medvedev, 562, L129 (2001).
\bibitem{Matt03} W. H. Matthaeus, G. Qin, J. W. Bieber, G. P. Zank, Astrophys. J., 590, L53 (2003).
\bibitem{Chan04} B. Chandran, J. Maron, Astrophys. J., 602, 170 (2004).
\bibitem{Maron04} J. Maron, B. Chandran, E. Blackman, PRL, 92, 045001 (2004).
\bibitem{kot00} J. K\'ota, J. R. Jokipii, ApJ, 531, 1067 (2000).
\bibitem{web06} G. M. Webb, G. P. Zank, E. Kh. Kaghashvili, J. A. le Roux, Astrophys. J., 651, 211 (2006).
\bibitem{ShaKo07a} A. Shalchi, I. Kourakis, Astronomy and Astrophysics, \textbf{470}, 405 (2007).
\bibitem{jok66} J. R. Jokipii, Astrophys. J., 146, 480 (1966).
\bibitem{mat95} W. H. Matthaeus, P. C. Gray, D. H. Jr. Pontius, J. W. Bieber, Phys. Rev. Lett., 75, 2136 (1995).
\bibitem{ShaKo07b} A. Shalchi, I. Kourakis, Physics of Plasmas, in press;
also as e-print astro-ph/0703366 at: http://arxiv.org/pdf/astro-ph/0703366
\bibitem{bie96} J. W. Bieber, W. Wanner, W. H. Matthaeus, J. Geophys. Res., \textbf{101}, 2511 (1996).
\bibitem{bru05} Bruno, R. \& Carbone, V., 2005, Living Reviews in Solar Physics, 2, 4
\bibitem{Grad} I. S. Gradshteyn, I. M. Ryzhik, {\it Table of integrals, series, and products}, (Academic Press, New York, 2000).
%
\end{thebibliography}
\end{document}